\documentclass[aps,prd,twocolumn,showpacs,10pt,superscriptaddress,preprintnumbers,nofootinbib]{revtex4-2}
\usepackage{amsmath,bm}
\allowdisplaybreaks[1]
\usepackage{physics}
\usepackage{graphicx}
\usepackage{amssymb}
\usepackage{xcolor}
\usepackage[colorlinks, linkcolor=blue!80!black, citecolor=green!50!black, urlcolor=magenta]{hyperref}
\usepackage[normalem]{ulem}
\usepackage{soul}
\usepackage{adjustbox}
\usepackage{orcidlink}

\usepackage{tikz}
\usetikzlibrary{arrows.meta}

\newcommand{\fig}[1]{Fig.~\ref{#1}}
\newcommand{\eq}[1]{Eq.~\eqref{#1}}
\newcommand{\eqs}[2]{Eqs.~\eqref{#1} and~\eqref{#2}}
\newcommand{\refcite}[1]{Ref.~\cite{#1}}
\newcommand{\refs}[1]{Refs.~\cite{#1}}

\newcommand{\pp}[1]{\left(#1\right)}
\newcommand{\bb}[1]{\left[#1\right]}

\newcommand{\vv}[1]{\left\langle #1 \right\rangle}
\newcommand{\bigpp}[1]{\big(#1\big)}

\newcommand{\beq}[1][]{\begin{equation}\label{#1}}
\newcommand{\eeq}{\end{equation}}
\newcommand{\bse}[1][]{\begin{subequations}\label{#1}}
\newcommand{\ese}{\end{subequations}}
\newcommand{\nn}{\nonumber}

\newcommand{\wt}[1]{\widetilde{#1}}

\newcommand{\M}{\mathcal{M}}

\newcommand{\F}{\mathcal{F}}
\renewcommand{\H}{\mathcal{H}}
\newcommand{\E}{\mathcal{E}}

\newcommand{\Ht}{\widetilde{\mathcal{H}}}
\newcommand{\Et}{\widetilde{\mathcal{E}}}

\def\GeV{{\rm GeV}}

\begin{document}

\preprint{
	{\vbox {			
		\hbox{\bf JLAB-THY-24-4181}
}}}
\vspace*{0.2cm}

\title{Unified and optimal frame choice for generalized parton distributions}

\author{Jian-Wei Qiu\,\orcidlink{0000-0002-7306-3307}}
\email{jqiu@jlab.org}
\affiliation{Theory Center, Jefferson Lab,
Newport News, Virginia 23606, USA}
\affiliation{Department of Physics, William \& Mary,
Williamsburg, Virginia 23187, USA}

\author{Nobuo Sato\,\orcidlink{0000-0002-1535-6208}}
\email{nsato@jlab.org}
\affiliation{Theory Center, Jefferson Lab,
Newport News, Virginia 23606, USA}

\author{Zhite Yu\,\orcidlink{0000-0003-1503-5364}}
\email{yuzhite@jlab.org (corresponding author)}
\affiliation{Theory Center, Jefferson Lab, 
Newport News, Virginia 23606, USA}

\date{\today}

\begin{abstract}
Reconstructing the internal three-dimensional quark and gluon structures of hadrons through generalized parton distributions (GPDs) from hard exclusive scattering processes is one of the most challenging tasks in nuclear and particle physics. In this paper, we introduce a new optimized reference frame that, for the first time, enables a unified view of all the reactions sensitive to GPDs and facilitates the interpretation of a variety of phase-space patterns that were previously hardly accessible and interpretable. Similarly to how the heliocentric description advanced our understanding of the solar system and gravitation, our new frame centers around a quasi-real state, allows for a consistent separation of physical scales, and reveals a novel quantum interference mechanism.
\end{abstract}

\maketitle

\section{Introduction}
Generalized parton distributions (GPDs) are fundamental quantum correlation functions (QCFs)
of quarks and gluons (collectively referred to as partons) 
inside hadrons and encode rich information that can unravel many aspects of the confined hadron structure, 
including quark and gluon tomography~\cite{Burkardt:2000za, Burkardt:2002hr}, 
mechanical properties~\cite{Polyakov:2002yz, Polyakov:2018zvc, Burkert:2018bqq}, 
and spin~\cite{Ji:1996ek} and mass~\cite{Ji:1994av, Ji:1995sv, Lorce:2017xzd, Metz:2020vxd} decompositions. 
Due to the color confinement of quantum chromodynamics (QCD), however, 
quarks, and gluons, and thus GPDs, cannot be measured {\it directly};
experiments only measure scattering cross sections of hadrons, leptons, and photons.
It is the QCD factorization formalisms~\cite{Collins:1996fb, Collins:1998be, Ji:1998xh, Qiu:2022pla}
that make it possible to extract GPDs {\it indirectly} from exclusive hadron scattering processes~\cite{Ji:1996nm, Radyushkin:1997ki, Brodsky:1994kf, Frankfurt:1995jw, Berger:2001xd, Guidal:2002kt, Belitsky:2002tf, Belitsky:2003fj, Kumano:2009he, Kumano:2009ky, Berger:2001zn, Goloskokov:2015zsa, Sawada:2016mao, ElBeiyad:2010pji, Pedrak:2017cpp, Pedrak:2020mfm, Siddikov:2022bku, Siddikov:2023qbd, Boussarie:2016qop, Duplancic:2018bum, Duplancic:2022ffo, Duplancic:2023kwe, Qiu:2022bpq, Qiu:2023mrm, Qiu:2024mny, Goeke:2001tz, Diehl:2003ny, Belitsky:2005qn, Boffi:2007yc}  
with controllable approximations, although the extraction is a challenging inverse problem.
Reconstructing GPDs from data has been and will continue to be 
a major research focus at the current and future high-energy nuclear
facilities~\cite{H1:2005gdw, H1:2009wnw, ZEUS:2008hcd, COMPASS:2018pup, CLAS:2001wjj, CLAS:2006krx, JeffersonLabHallA:2006prd, CLAS:2007clm, CLAS:2008ahu, CLAS:2015bqi, JeffersonLabHallA:2015dwe, CLAS:2015uuo, Defurne:2017paw, CLAS:2017udk, Benali:2020vma, CLAS:2021ovm, JeffersonLabHallA:2022pnx, CLAS:2022syx, Accardi:2012qut, AbdulKhalek:2021gbh, Anderle:2021wcy, Accardi:2023chb}.

One of the key challenges is that the measurable cross sections generally receive mixed contributions from different types of QCFs.
Recognizing the fact that the azimuthal distribution of the observed final-state particle(s)
is directly sensitive to the characteristics of individual subprocesses, such modulations 
have been used to separate contributions from different QCFs.  
For example, it has been used for disentangling contributions from different transverse momentum-dependent parton distributions (TMDs) 
in lepton-hadron semi-inclusive deep inelastic scattering (SIDIS), 
$e(\ell)+h(p)\to e(\ell')+h'(p')+X$ 
by measuring azimuthal modulations of the angle between the leptonic plane of
$e(\ell)\to e(\ell')$
and the hadronic plane of 
$h(p)\to h'(p')$~\cite{Diehl:2005pc, Bacchetta:2006tn, Bacchetta:2008xw}. 

However, when extracting GPDs from {\it exclusive} lepton-hadron processes, 
e.g., $e(\ell)+h(p)\to e(\ell')+h'(p')+\gamma(p_\gamma)$, 
the angular distributions between the same leptonic and hadronic planes 
are sensitive to not only the contributions from different GPDs through the deeply virtual Compton scattering (DVCS) subprocess, 
but also the contributions from the subprocess not even associated with GPDs,
known as the Bethe-Heitler (BH) subprocess, which makes
the separation of different GPDs a very challenging and long-standing issue since the 1990's~\cite{Belitsky:2001ns, Guidal:2008ie, Guidal:2009aa, Guidal:2010ig, Guidal:2010de, Guidal:2013rya, Boer:2014kya, Kumericki:2013br, Kumericki:2007sa, Kumericki:2009uq, Kumericki:2011rz, Goldstein:2010gu, Kumericki:2016ehc, Kriesten:2019jep, Kriesten:2020wcx, Kriesten:2020apm, Almaeen:2022imx, Almaeen:2024guo, Guo:2021gru, Shiells:2021xqo, Grigsby:2020auv, Moutarde:2019tqa, Cuic:2020iwt, Berger:2001xd}.
This issue becomes increasingly challenging when considering higher-twist contributions~\cite{Guo:2021gru, Bhattacharya:2022aob, Bhattacharya:2023jsc}. 

In this paper, we show that all these difficulties are rooted in the choice of frame for the study of GPD-sensitive exclusive reactions,
\beq[eq:sdhep]
	h(p) + B(p_2) \to h'(p') + C(q_1) + D(q_2)\, ,
\eeq
where a hadron $h$ is diffracted into $h'$ by an incident beam particle $B \, (=e,\gamma,\pi,...)$ with a soft diffractive momentum scale $t=(p-p')^2$ 
and produces a pair of final-state particles $C$ and $D$ with large balancing transverse momenta with respect to the $h$-$B$ collision axis, 
$q_{1T} \sim q_{2T} \sim q_T\gg \sqrt{-t}$. 
Under these kinematic conditions, 
\eq{eq:sdhep} unifies all $2\to3$ GPD-sensitive processes,
which we refer to as {\it single-diffractive hard exclusive processes} (SDHEPs)~\cite{Qiu:2022pla}.
The DVCS is only a subprocess of the 
specific SDHEP with $B=C=e$ and $D=\gamma$.

While an SDHEP can, in principle, be described in different frames, 
the features of produced particle distributions can be better represented in one frame than in others. 
By analogy to the geocentric and heliocentric descriptions of the solar system, adopting an optimal frame not only simplifies the kinematics but also influences how the dynamics are conceptualized and formulated.

With $q_T\gg \sqrt{-t}$, an SDHEP is naturally
described in two stages,
\bse\label{eq:two-stage}
\begin{align}
    & h(p) \to A^*(\Delta = p - p') + h'(p'), 	\label{eq:diffractive}\\
    & \hspace{8.5ex} \begin{tikzpicture}
        \node[inner sep=0pt] (arrow) at (0, 0) {
            \tikz{\draw[->, >={Stealth}, double, double distance=1pt, line width=1pt] (0, 0.28) to [out=-90, in=180] (0.68, 0);}
        };
    \end{tikzpicture} 
    \hspace{0.5ex}
    A^*(\Delta) + B(p_2) \to C(q_1) + D(q_2),
    \label{eq:hard 2to2}
\end{align}\ese
at a soft scale $\sqrt{-t}$ and a hard scale $q_T$, respectively, linked by a long-lived virtual state $A^*$ of momentum $\Delta$.
This $A^*$ state defines the quantum exchange between the diffractive hadron in \eq{eq:diffractive}
and the hard exclusive collision in \eq{eq:hard 2to2}. Thus, naturally centering the kinematics of the reaction around $A^*$ gives a clear formulation of the whole SDHEP amplitude.
The BH subprocess corresponds to $A^* = \gamma^*$, contributing at leading power (LP) in $\sqrt{-t}/q_T$, 
while the GPD-sensitive subprocess occurs at the next-to-leading power (NLP) with 
$A^*$ being a pair of quark and antiquark $[q\bar{q}]$ or gluons $[gg]$.
Three- or more-parton channels of $A^*$ contribute at even higher powers.  
To separate these different subprocesses, we introduce a new SDHEP frame applicable to all the reactions in \eq{eq:sdhep}
so that the azimuthal structure of produced particles directly reflects the spin and parity property of the exchange-state $A^*$.

\begin{figure}[htp]
    \centering
    \includegraphics[scale=0.45]{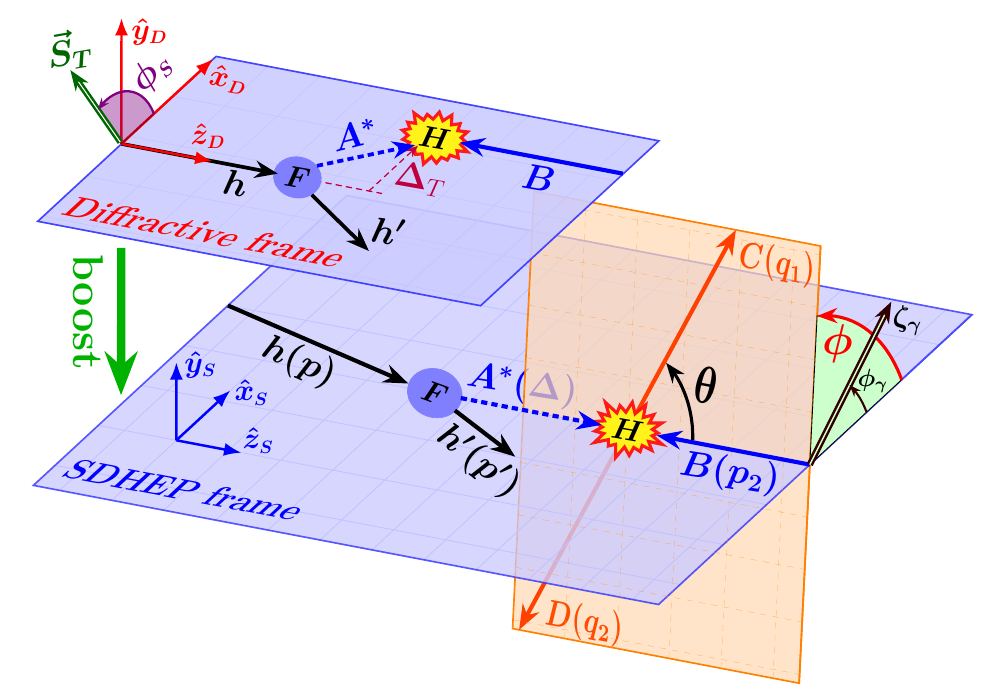} 
	\caption{Frames for analyzing SDHEPs in \eq{eq:sdhep}.
        The big vertical arrow refers to the Lorentz transformation from the diffractive frame to the SDHEP frame. The linear polarization $\zeta_{\gamma}$ along $\phi_{\gamma}$ applies only to photoproduction processes.}
	\label{fig:frame}
\end{figure}

\section{Azimuthal modulations in SDHEPs}
The {\it SDHEP frame} is defined as the center-of-mass (c.m.) frame of the $2\to 2$ scattering  in \eq{eq:hard 2to2},
with the $A^*$ moving along the $\hat{z}_S$-axis and the $\hat{x}_S$-axis lying in the {\it diffractive plane} for $h(p)\to h'(p')$ along the opposite direction of the transverse momentum $\bm{p}_T$ of $h$, 
as shown in the lower part of \fig{fig:frame}. 
The produced particles $C$ and $D$ define the {\it scattering plane} that intersects with the diffractive plane along the $\hat{z}_S$ axis, making an angle $\phi$.  
The advantage of the SDHEP frame is that the $\hat{z}_S$ axis is along the collision of two real or quasireal particles such that 
the distribution of the azimuthal angle $\phi$ is uniquely determined by the spin states of $A^*$ and $B$.

The hadron momentum $p$ and $p'$ in the SDHEP frame can be obtained from a Lorentz transformation from the {\it diffractive frame}, 
which is the c.m.\ of the collision between hadron $h$ and beam particle $B$, 
with the $\hat{z}_D$ axis oriented along the direction of $h$ and $\hat{x}_D$ axis along $\bm{\Delta}_T \equiv (\bm{p} - \bm{p}')_T$, as shown in the upper part of \fig{fig:frame}. 
Notably, the $\hat{x}_D$ axis varies event by event, and trades the azimuthal angle $\phi_{\Delta}$ of $\bm{\Delta}_T$ with respect to a fixed lab coordinate system
for the azimuthal angle $\phi_S$ of the hadron's transverse spin $\bm{S}_T$, 
similarly to the treatment of SIDIS~\cite{Diehl:2005pc, Bacchetta:2004jz, Bacchetta:2006tn, Bacchetta:2008xw}. 
The distribution of $\phi_{\Delta}$ (and thus of $\phi_S$) is nontrivial only when $S_T\equiv|\bm{S}_T| \neq 0$, 
allowing us to characterize the diffraction process with variables $\phi_S$, $t$, and the skewness  $\xi = [(p - p') \cdot n] / [(p + p') \cdot n]$, 
where the lightlike vector $n^{\mu}$ is along the direction of $B(p_2)$. 

We note that the transformation from the diffractive frame to the SDHEP frame is technically a transverse boost~\cite{Diehl:2003ny} (up to corrections of order $M_B^2/q_T^2$ with $M_B$ being the $B$ mass, which vanishes for lepton and photon beams), which keeps the vector $n$, and thus the $\xi$ and GPDs, unchanged. 
The kinematics of SDHEP events is then completely described in terms of five variables $(t, \xi, \phi_S, \theta, \phi)$.

With this setup, the total scattering amplitude of the reaction in Eq.~(\ref{eq:two-stage})
can be written schematically as
\begin{align}
    \M^{hB\to h'CD}_{\lambda_h \lambda_B} &= \sum_{A^*} 
    e^{-i \lambda_h \phi_S} F^{h \to h'A^*}_{\lambda_h}(t, \xi)    \nn\\ 
    &\hspace{3em} \otimes 
    e^{i\pp{\lambda_{A} - \lambda_B} \phi}
    H_{\lambda_A \lambda_B}^{A^*B \to CD}(\hat{s}, \theta),
\label{eq:sdhep-M-decomp-A-phi}
\end{align}
where the two stages for each $A^*$ channel are factorized in terms of the diffractive structure function $F^{h \to h'A^*}_{\lambda_h}$ and hard scattering coefficient $H_{\lambda_A \lambda_B}^{A^*B \to CD}$. Importantly, two distinct azimuthal phases emerge, one from each frame: a $\phi_S$ dependence in the diffractive frame controlled by the initial hadron helicity $\lambda_h$, and a $\phi$ dependence in the SDHEP frame associated with the hard scattering, determined by the helicities $\lambda_A$ and $\lambda_B$ of the $A^*$ and $B$ particles, respectively.
In general, when computing physical observables, one needs to square Eq.~(\ref{eq:sdhep-M-decomp-A-phi}) and trace over the spin density matrices for the incoming particles, i.e., 
$ \rho^{(h)}_{\lambda_h\lambda'_h} \rho^{(B)}_{\lambda_B\lambda'_B} \M_{\lambda_h \lambda_B} \M_{\lambda'_h \lambda'_B}^*$, 
which causes different $h$ and $B$ helicities as well as different $A^*$ channels to interfere, 
giving rise to a variety of azimuthal modulations in $\phi_S$ and $\phi$. 
For instance, for a nucleon target with transverse spin $S_T$, the interference of $\lambda_h = \pm 1/2$ leads to $\cos\phi_S$ and $\sin\phi_S$ modulations,
while the interference between two $(A^*, B)$ channels of helicities $(\lambda_A, \lambda_B)$ and $(\lambda_A^{\prime}, \lambda'_B)$ would lead to the azimuthal modulations 
$\cos[ (\Delta\lambda_A - \Delta\lambda_B) \phi ]$ and $\sin[ (\Delta\lambda_A - \Delta\lambda_B) \phi ]$,
with $(\Delta\lambda_A, \Delta\lambda_B) \equiv (\lambda_A - \lambda_A', \lambda_B - \lambda'_B)$.  

It is these azimuthal modulations in $\phi$ and $\phi_S$ that ultimately allow us to separate different GPDs.
Unlike those in SIDIS, the modulations in SDHEPs contain unique effects stemming from the interference between particles of different species and numbers, characteristic of multiparticle interference featured in high-twist inclusive processes~\cite{Qiu:1991pp}. 
With the QCD factorization at the amplitude level for exclusive processes, 
it is the choice of the SDHEP frame that makes such interference pattern coherently formulated for the first time
and applicable to all $2\to3$ exclusive processes for extracting GPDs.

\section{Electroproduction processes}
A classical example is the production of a real photon off a nucleon $N$, i.e., 
$h = h' = N$, $B = C = e$, and $D = \gamma$ in \eqs{eq:sdhep}{eq:two-stage}.

\begin{figure}[htbp]
	\centering
	\begin{tabular}{cc}
		\includegraphics[scale=0.5]{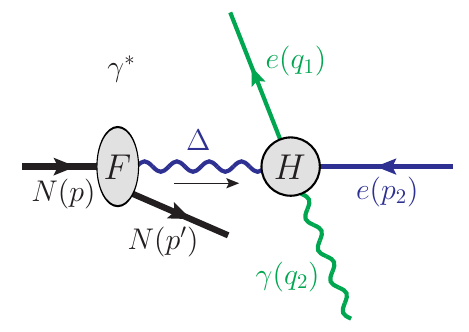} &
		\includegraphics[scale=0.5]{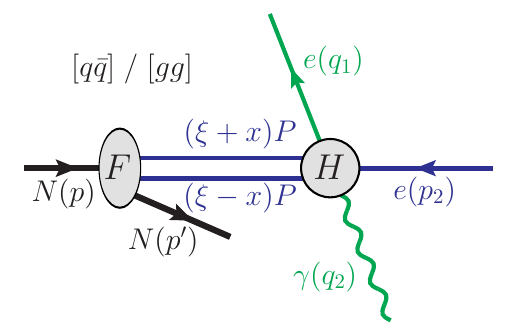} \\
		(a) & (b)
	\end{tabular}
	\caption{The (a) leading and (b) subleading channels in the photon electroproduction.
        The two parton lines in (b) are either quarks or gluons.}
	\label{fig:dvcs}
\end{figure}

The $A^*$ can in general be a state with $j\ge 1$ particles.
In a physical gauge, the amplitude with $(j+1)$ particles in $A^*$ is suppressed by a factor of $\sqrt{-t}/q_T$ relative to the amplitude with $j$ particles in $A^*$. 
The leading channel is the BH subprocess in \fig{fig:dvcs}(a) with $A^* = \gamma^*$ for $j = 1$, 
which has three possible helicity states, $\lambda_A^{\gamma} = \pm 1$ or $0$.
The amplitude of this subprocess can be written as
\begin{align}
	&\M^{[1]} 
	 	= \frac{-e}{t} \, F_N^{\mu}(p, p') \, G^{\gamma}_{\mu}(\Delta, p_2, q_1, q_2)	\nn\\
	&\hspace{0em} 
		= \frac{e}{t} \bb{ 
			\sum_{\lambda = \pm1} (F_N \cdot \epsilon^*_{\lambda}) (\epsilon_{\lambda} \cdot G^{\gamma})
			- 2 (F_N \cdot n) (\bar{n} \cdot G^{\gamma} )
		},
\label{eq:dvcs-a}
\end{align}
where $F_N^{\mu}(p, p') = \vv{ N(p') | J^{\mu}(0) | N(p) }$ is parametrized by the standard nucleon electromagnetic form factors $F_1$ and $F_2$.
The hard coefficient $G^{\gamma}_{\mu}$ describes the scattering of the virtual photon with the electron.
In the second line of \eq{eq:dvcs-a}, we use the kinematics in the SDHEP frame to decompose the photon propagator into a sum of three polarizations, with the polarization vector being 
$\epsilon_{\pm}^{\mu} = (0, \mp 1, - i, 0) / \sqrt{2}$
for the transverse $\gamma_T^*$ and 
$\epsilon_0^{\mu} = \bar{n}^{\mu} = (1, 0, 0, 1) / \sqrt{2}$ 
for the longitudinal $\gamma_L^*$. 
One can show by Ward identity that only the $\gamma_T^*$ contributes to the amplitude at LP, while the $\gamma_L^*$ amplitude is NLP, 
which is at the same power as the two-parton channels ($j = 2$) in \fig{fig:dvcs}(b), $A^* = [q\bar{q}]$ or $[gg]$.
The amplitude of the latter can be factorized into GPDs $F$ and $\wt{F}$, with perturbative coefficients $G$ and $\wt{G}$, respectively,
\begin{align}
	\M^{[2]}
		&= \sum_q \int_{-1}^1 dx \big[ F^q(x, \xi, t) \, G^q(x, \xi; \hat{s}, \theta, \phi)	\nn\\
		&
			+ \wt{F}^q(x, \xi, t) \, \wt{G}^q(x, \xi; \hat{s}, \theta, \phi) \big] + \order{\sqrt{-t} / q_T},
\label{eq:dvcs-factorize-M}
\end{align}
where we have suppressed the factorization scale and neglected the gluon GPD contributions. 
The amplitudes $\M^{[\geq 3]}$ involve three- or more-parton channels of $A^*$ entering at 
next-to-next-to-leading power (NNLP) or beyond, together with the high-twist effects in \eq{eq:dvcs-factorize-M}~\cite{Anikin:2000em, Penttinen:2000dg, Belitsky:2000vx, Kivel:2000cn, Radyushkin:2000ap, Diehl:2003ny, Belitsky:2001ns, Belitsky:2005qn, Kriesten:2019jep, Guo:2022cgq, Kivel:2000rb, Radyushkin:2000jy, Radyushkin:2001fc, Aslan:2018zzk, Blumlein:2006ia, Blumlein:2009hit, Braun:2011zr, Braun:2011dg, Braun:2012bg, Braun:2012hq, Braun:2014sta}.  

Hence, the full amplitude $\M$  receives a LP contribution from $\gamma_T^*$, with helicities $\lambda_A^{\gamma} = \pm 1$,
and NLP corrections from the $\gamma_L^*$ and $[q\bar{q}]$ channels, both with helicities $0$. Importantly, the $\phi$ dependence of the corresponding hard coefficients in \eqs{eq:dvcs-a}{eq:dvcs-factorize-M} are given by $\epsilon_{\pm} \cdot G^{\gamma} \propto e^{\pm i \phi}$ and  $(\bar{n} \cdot G^{\gamma}, G^q, \wt{G}^q) \propto e^{0 \cdot i \phi}$, where we suppress the $\phi$ dependence stemming from the electron helicity since it disappears in the amplitude squared. 
Therefore, up to the NLP accuracy, the $|\M|^2$ includes the $\gamma_T^*$ amplitude squared, which gives a flat $\phi$ distribution, 
and its interference with the $\gamma_L^*$ and $[q\bar{q}]$ amplitudes, which gives rise to $\cos\phi$ and $\sin\phi$ modulations. 
The resulting NLP cross section for the photon electroproduction process is
\begin{align}
	&\frac{d\sigma^{\gamma}_e}{dt \, d\xi \, d\phi_S \, d\cos\theta \, d\phi}
	= \frac{d\sigma^{\gamma, {\rm unp.}}_e}{dt \, d\xi \, d\cos\theta} \,
    \Omega^{\gamma}_e(\phi_S, \phi),
\label{eq:dvcs-xsec}
\end{align}
where $d\sigma^{\gamma, {\rm unp.}}_e$ is the unpolarized differential cross section and $\Omega^{\gamma}_e(\phi_S, \phi)$ modulates the azimuthal distributions as
\begin{widetext}
\begin{align}
	(2\pi)^2 \Omega^{\gamma}_e(\phi_S, \phi)
	&= 1 + P_e P_N A^{\rm LP}_{LL} + P_e S_T A^{\rm LP}_{TL}  \cos\phi_S
	+ \pp{ A_{UU}^{\rm NLP} + P_e P_N A_{LL}^{\rm NLP} } \cos\phi 
			+ \pp{ P_e A_{UL}^{\rm NLP} + P_N A_{LU}^{\rm NLP} }\sin\phi 	\nn\\
	& + S_T \pp{ A_{TU, 1}^{\rm NLP} \cos\phi_S \sin\phi + A_{TU, 2}^{\rm NLP} \sin\phi_S \cos\phi }	
			+ P_e S_T \pp{ A_{TL, 1}^{\rm NLP} \cos\phi_S \cos\phi + A_{TL, 2}^{\rm NLP} \sin\phi_S \sin\phi},
\label{eq:dvcs-phi}
\end{align}
\end{widetext}
where $P_e$ and $P_N$ are the net helicities of the electron and nucleon beams, respectively. The polarization asymmetries $A$'s given in terms of form factors $F_{1,2}$ and GPDs can be found in \cite{supplemental}. Their  superscripts refer to the power of $\sqrt{-t}/q_T$ at which they contribute and their subscripts refer sequentially to the nucleon and electron polarizations, with ``$U$'', ``$L$'', and ``$T$'' standing for ``unpolarized'', ``longitudinally polarized'', and ``transversely polarized'', respectively. As discussed, the $\cos\phi_S$ and $\sin\phi_S$ modulations arise only when the nucleon has a nonzero transverse spin $S_T$. 
At NLP, both modulations are present in $(TU)$ and $(TL)$ configurations, and we distinguish them with additional subscripts $1,2$. 

The LP asymmetries only depend on the form factors, $F_1$ and $F_2$, 
whereas the NLP asymmetries depend on the GPD moments~\cite{Qiu:2023mrm, supplemental} in a linear form. 
The $\cos\phi$ and $\sin\phi$ modulations are unambiguous signatures of GPDs in the photon electroproduction process, even without knowledge of GPDs.
Up to this order, we have eight NLP asymmetries in total, corresponding exactly to the eight real degrees of freedom of the GPD moments. 
A complete measurement of all the polarization asymmetries, extracted with the aid of the azimuthal projections, can disentangle all the GPD moments.

The azimuthal modulation analysis is based on rotational symmetry which extends to all orders in the strong coupling in perturbative QCD for the  helicity-zero GPDs $F^{q, g}$ and $\wt{F}^{q, g}$. 
Beyond leading order, the photon electroproduction at NLP 
receives contributions from the gluon transversity GPD $F^g_T$, which carries a helicity $\lambda_A^{g_T} = \pm 2$ and 
interferes with the $\gamma_T^*$ amplitude to give  $\cos3\phi$ and $\sin3\phi$ modulations.
The NNLP contributions include the square of the $\gamma_L^*$ and $[q\bar{q}]$ amplitudes and the interference of the $\gamma_T^*$ with three-parton channels, $A^* = [q\bar{q}g]$ or $[ggg]$. The latter carry helicities $\lambda_A^{qqg} = \pm 1$ and $\lambda_A^{ggg} = \pm 1$ or $\pm 3$. 
These give rise to additional azimuthal modulations $(\cos2\phi, \, \sin2\phi)$ as well as $(\cos4\phi, \, \sin4\phi)$. 
This pattern builds up and generates a tower of azimuthal modulations with inclusion of higher twists.

\begin{figure}[htbp]
    \centering
	\includegraphics[trim={1em 1em 7em 2em}, clip, scale=0.23]{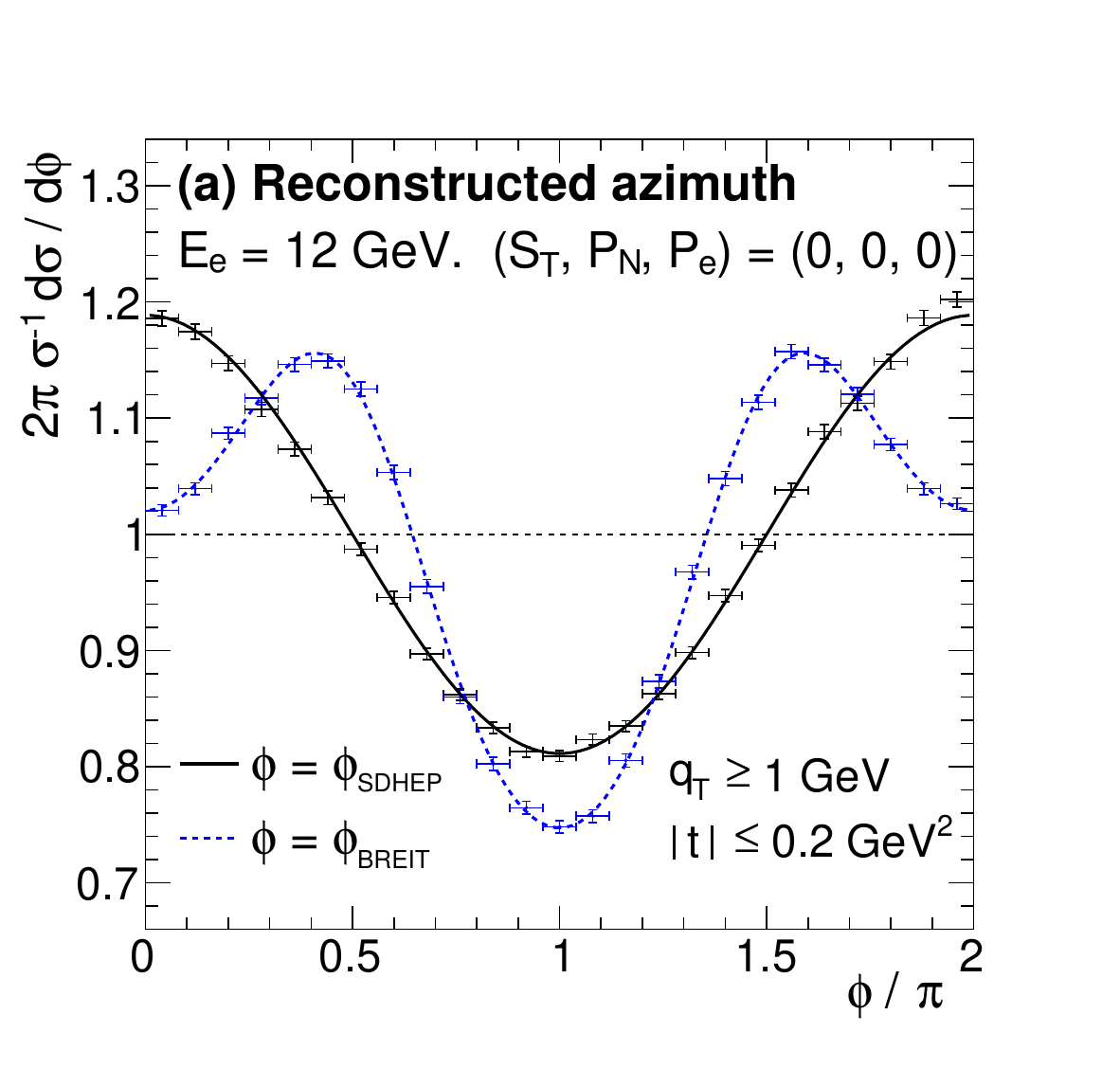} \;\;
        \includegraphics[trim={1em 1em 7em 2em}, clip, scale=0.23]{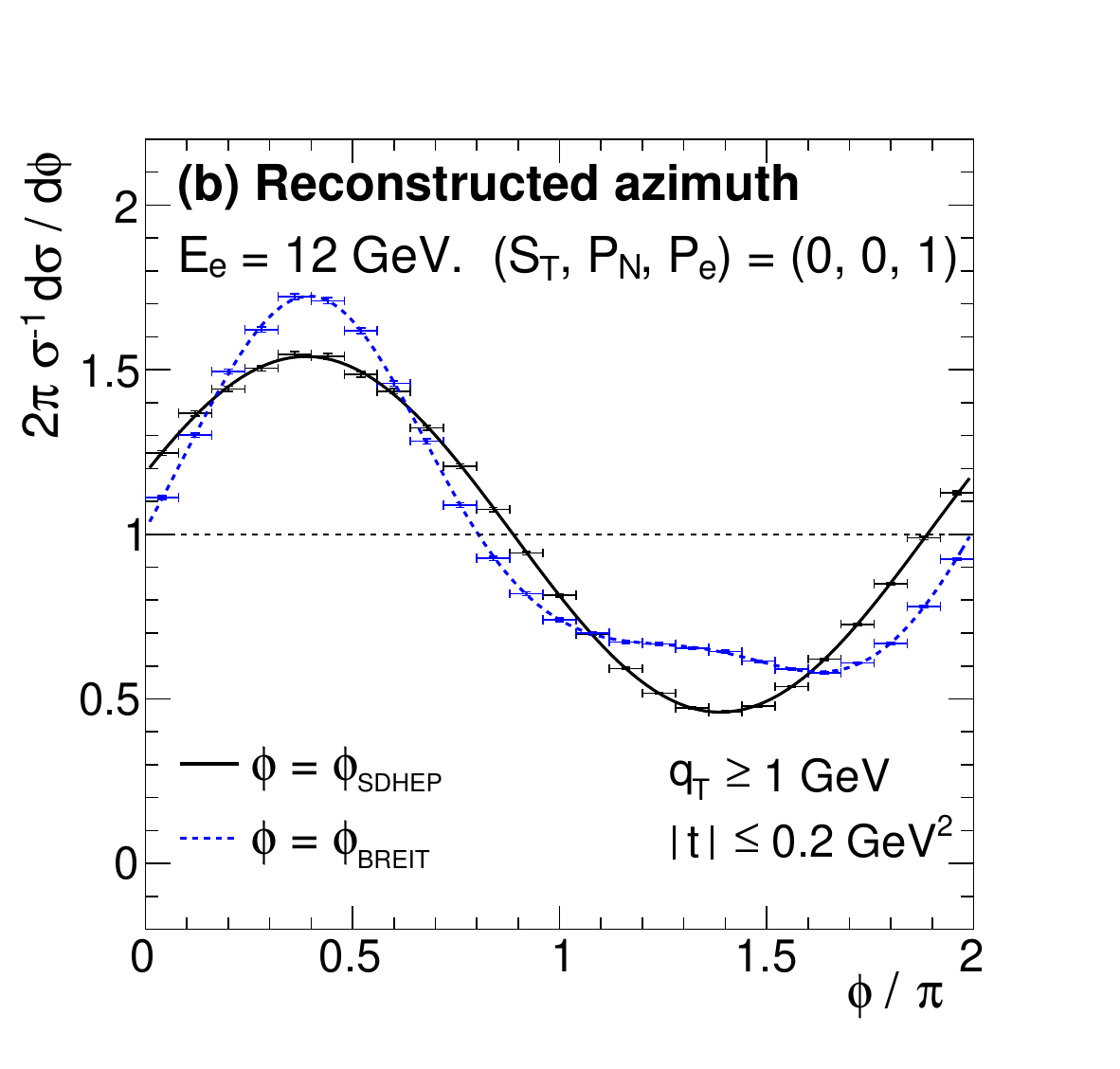} 
    \caption{Monte Carlo reconstructed azimuthal $\phi$ distributions for (a) unpolarized case and (b) single electron polarization. 
    The two curves in each figure refer to the azimuthal distributions reconstructed in the SDHEP and Breit frames.}
    \label{fig:phi}
\end{figure}

Our formalism contrasts with the traditional approach, which is based on the angular distribution between the leptonic and hadronic planes in the Breit frame, whose $z$-axis is along the collision between 
the hadron $h(p)$ and the exchanged virtual photon $\gamma_{ee}^*(q=p_2-q_1)$ of the DVCS subprocess. 
The BH subprocess contributing to the same cross section does not have the same state of momentum $q$ and
makes the azimuthal modulations highly nontrivial and difficult to interpret~\cite{Belitsky:2001ns}.
In \fig{fig:phi}(a) we show the unpolarized $\phi$ distributions evaluated using the nucleon form factors from \refcite{Kelly:2004hm} 
and GPD models from \refs{Goloskokov:2005sd, Goloskokov:2007nt, Goloskokov:2009ia, Kroll:2012sm} in the SDHEP (solid line) and Breit (dotted line) frames, respectively.  
We performed the simulation with an electron beam of energy $E_{e} = 12~\GeV$, 
restricting the final state phase space to $q_T \geq 1~\GeV$ and $|t| \leq 0.2~\GeV^2$.
The SDHEP-based $\phi$ modulation exhibits a clear $\cos\phi$ structure with an amplitude that is exactly $A_{UU}^{\rm NLP}$, as shown in \eq{eq:dvcs-phi}, 
while the Breit-frame modulation can be fit by a harmonic series $\sum^N_{n=0} a_n \cos n\phi$ with at least $N = 4$.
Due to the presence of $\phi$ dependence in the denominators, 
the Breit-frame distribution introduces additional systematic uncertainties 
when relating measurements to the underlying GPDs~\cite{Belitsky:2001ns, Shiells:2021xqo, Kriesten:2019jep}.
Similarly, \fig{fig:phi}(b) shows the azimuthal distributions with a polarized electron beam $(P_e = 1)$. 
The SDHEP-frame distribution (solid line) contains an additional $\sin\phi$ component on top of the unpolarized distribution with an amplitude exactly equal to $A_{UL}^{\rm NLP}$. 
In contrast, the Breit-frame distribution (dotted line) exhibits a highly nontrivial azimuthal modulation;
the obvious need to require a ``proper'' fitting template adds significant ambiguity to the extraction of the GPD signals.

In our two-stage kinematic description, the $\phi_S$ modulations are also regular and predictable,
further facilitating the extraction of GPDs, 
while the $\phi_S$ distribution in the Breit frame is distorted 
due to a Jacobian factor from the lab-to-Breit frame transformation~\cite{Diehl:2005pc}. 
These effects are power suppressed in SIDIS~\cite{Bacchetta:2006tn, Bacchetta:2008xw}, 
but can become important in reconstructing GPDs because the latter themselves enter the observables at NLP. 
However, these effects are typically ignored in the literature~\cite{Belitsky:2001ns}.

Like the photon electroproduction, meson electroproduction process also includes the $A^* = \gamma^*$ channel~\cite{Qiu:2022pla}. 
However, this channel is electromagnetically suppressed relative to the two-parton channels $A^* = [q\bar{q}]$/$[gg]$ and can be neglected,
and the Breit-frame azimuthal distribution is relatively regular and less critically dependent on the choice of frame. 
However, GPDs are not the same in different frames. 
Although the frame-induced difference is power suppressed and could be potentially addressed by a full high-twist analysis~\cite{Guo:2021gru}, it may not be numerically small at current energies of experiments. It is also unclear whether an  all-order twist-3 factorization holds~\cite{Duplancic:2023xrt}.
Therefore, analyzing both the photon and meson electroproduction processes in the SDHEP frame grants us the same $n^{\mu}$ and GPD definitions.

\section{Photoproduction processes}
Let us first consider the dilepton photoproduction off a nucleon, i.e., $h = h' = N$, $B = \gamma$, and $(C,D) = (\ell^-, \ell^+)$, as an example in this category.  
This reaction contains both the BH channel with $A^* = \gamma^*$ and two-parton channels with $A^* = [q\bar{q}]$ or $[gg]$, with the hard-scattering subprocesses  
$\gamma^* + \gamma \to \ell^- + \ell^+$ and 
$[q\bar{q}] / [gg] + \gamma \to \gamma^*_{\ell\ell} \to \ell^- + \ell^+$, 
respectively. The main difference from the electroproduction is that the photon beam can carry a linear polarization $\zeta_{\gamma}$ in addition to a net helicity (circular polarization) $P_{\gamma}$, which allows the interference of opposite photon beam helicities. This injects two additional units of helicity flips, $\Delta\lambda_B = \pm 2$,
and gives rise to $\cos3\phi$ and $\sin3\phi$ modulations in the interference of one- and two-parton channels.

Up to NLP, the cross section $d\sigma^{\ell\ell}_{\gamma}$ takes a form similar to \eq{eq:dvcs-xsec}, with the azimuthal modulation given by
\begin{widetext}
\begin{align}
	(2\pi)^2 \Omega^{\ell\ell}_{\gamma}(\phi_S, \phi)
	&= 1 + P_N P_{\gamma} A^{\rm LP}_{LL} 
			+ S_T P_{\gamma} A^{\rm LP}_{TL} \cos\phi_S 
			+ \zeta_{\gamma} A^{\rm LP}_{UT} \cos(4\phi - 2 \phi_{\gamma})
			+ \pp{ A_{UU}^{\rm NLP} + P_N P_{\gamma} A_{LL}^{\rm NLP} } \cos\phi 
		\nn\\
		&\hspace{1.82em}
			+ \pp{ P_N A_{LU}^{\rm NLP} + P_{\gamma} A_{UL}^{\rm NLP} }\sin\phi	
			+ \zeta_{\gamma} A_{UT}^{\rm NLP} \cos(3\phi - 2\phi_{\gamma})
			+ P_N \zeta_{\gamma} A_{LT}^{\rm NLP} \sin(3\phi - 2\phi_{\gamma})  	\nn\\
		&\hspace{1.82em}
			+ S_T \pp{ A_{TU, 1}^{\rm NLP} \cos\phi_S \sin\phi + A_{TU, 2}^{\rm NLP} \sin\phi_S \cos\phi }
			+ S_T P_{\gamma} \pp{ A_{TL, 1}^{\rm NLP} \cos\phi_S \cos\phi + A_{TL, 2}^{\rm NLP} \sin\phi_S \sin\phi }		\nn\\
		&\hspace{1.82em}
 			+ S_T \zeta_{\gamma} \bb{ A_{TT, 1}^{\rm NLP} \cos\phi_S \sin(3\phi - 2\phi_{\gamma}) + A_{TT, 2}^{\rm NLP} \sin\phi_S \cos(3\phi - 2\phi_{\gamma}) }\;.
\label{eq:tcs-phi}
\end{align}
\end{widetext}
Here, $\phi_{\gamma}$ refers to the direction of the linear photon polarization, as shown in \fig{fig:frame}, and the expressions for the asymmetries can be found in \cite{supplemental}. 
Similar to \eq{eq:dvcs-phi}, the LP asymmetries only depend on $F_{1,2}$  while the NLP asymmetries are linear in the GPD moments~\cite{Berger:2001xd}. 

Differently from the photon electroproduction, the azimuthal dependence in \eq{eq:tcs-phi} occurs at LP 
because the linear polarization $\zeta_{\gamma}$ of the incoming photon causes an interference between the two $\gamma_T^*$ 
helicity states, leading to a $\cos(4\phi - 2\phi_{\gamma})$ modulation. 
Also, four additional asymmetries $(A_{UT}^{\rm NLP}, A_{LT}^{\rm NLP}, A_{TT, 1}^{\rm NLP}, A_{TT, 2}^{\rm NLP})$ are present and 
all induced by $\zeta_{\gamma}$. 
These are not independent but provide valuable constraints for reconstructing GPDs, especially when transversely polarized targets are unavailable.

We note that the chosen frame to study this process in the literature~\cite{Berger:2001xd, Boer:2015fwa, Boer:2015cwa, Nadel-Turonski:2009gdz, CLAS:2021lky} is similar to our SDHEP frame, 
but with the $\hat{z}$ axis along the opposite direction of the diffracted nucleon, i.e., $-\bm{p}'$. 
Consequently, the $\phi$ definition changes in a nontrivial manner and mixes the soft and hard dynamics in the BH channel,
leading to additional $\phi$-dependent factors that distort the regularity of the azimuthal modulation
in a way similar to the Breit-frame description of photon electroproduction. 
We stress that in the presence of the $A^*=\gamma^*$ subprocess, the only way to avoid irregularities in azimuthal modulations is to analyze the process in the SDHEP frame.

The same analysis can be applied to other photoproduction processes, including the diphoton production~\cite{Pedrak:2017cpp, Grocholski:2021man, Grocholski:2022rqj}, photon-meson pair production~\cite{Boussarie:2016qop, Duplancic:2018bum, Duplancic:2022ffo, Duplancic:2023kwe, Qiu:2022pla, Qiu:2023mrm}, and meson-pair production~\cite{ElBeiyad:2010pji, Siddikov:2022bku, Siddikov:2023qbd}. 

\section{Mesoproduction processes}
Dilepton~\cite{Berger:2001zn, Goloskokov:2015zsa, Sawada:2016mao} or diphoton~\cite{Qiu:2022bpq, Qiu:2024mny} mesoproduction processes can be studied within the SDHEP framework. 
Since only charged scalar meson beams like $\pi^{\pm}$ or $K^{\pm}$ are available,
the $A^*=\gamma^*$ and $[gg]$ channels are not allowed. Consequently, the
only possible $\phi$ modulations arise from the interference between twist-2 and higher-twist transition GPDs, 
making the SDHEP frame the most natural choice.

\section{Summary and outlook}
Centering the kinematics around the intermediate quasireal state $A^*$,
the two-stage SDHEP framework provides a unified description for all $2\to3$ GPD-sensitive exclusive processes
with a consistent separation of dynamics at two different momentum scales.
The new formalism elucidates the quantum interference mechanism responsible for the various azimuthal modulations and 
produces a regular harmonic series that is readily suited for disentangling and extracting different types of GPDs.
The full phenomenological realization of GPD extraction from existing experiments and the future Electron-Ion Collider, using our developed SDHEP framework, 
is key to uncovering the femtoscopic details of hadronic and nuclear systems that emerge from QCD.

\vspace{3mm}
We thank Y.~Guo, E.~Moffat, and M.~G.~Santiago for helpful discussions and communications. 
This work was supported in part by the U.S. Department of Energy (DOE) Contract No.~DE-AC05-06OR23177, 
under which Jefferson Science Associates, LLC operates Jefferson Lab, the Jefferson Lab LDRD program Projects No.~LD2312 and No.~LD2406. 
The work of N.S. was also supported by the DOE, Office of Science, Office of Nuclear Physics in the Early Career Program.

\bibliographystyle{apsrev}
\bibliography{reference}


\clearpage
\onecolumngrid

\begin{center}
    \textbf{\large Supplemental Material for \\ \vspace{0.1cm} ``Unified and optimal frame choice for generalized parton distributions''}
    
    \vspace{0.4cm}
    Jian-Wei Qiu,$^{1,2}$ Nobuo Sato,$^1$ and Zhite Yu$^1$
    
    \vspace{0.2cm}
    \textit{\small $^1$Theory Center, Jefferson Lab, Newport News, Virginia 23606, USA \\
           $^2$Department of Physics, William \& Mary, Williamsburg, Virginia 23187, USA}
           
    {\small (Dated: \today)}
\end{center}

\vspace{-0.2in}


\makeatletter
\renewcommand\p@subsection{}
\makeatother

\setcounter{equation}{1000}
\setcounter{figure}{1000}
\setcounter{table}{1000}

\renewcommand{\theequation}{S\the\numexpr\value{equation}-1000\relax}
\renewcommand{\thefigure}{S\the\numexpr\value{figure}-1000\relax}
\renewcommand{\thetable}{S\the\numexpr\value{table}-1000\relax}

\subsection*{A:\ Polarization asymmetry coefficients for the real photon electroproduction process}
\label{app:dvcs}
\vspace{-0.1in}
We first denote the unpolarized cross section in \eq{eq:dvcs-xsec} and the polarization asymmetry coefficients in \eq{eq:dvcs-phi} as
\beq
	\frac{d\sigma_e^{\gamma, {\rm unp.}}}{dt \, d\xi \, d\cos\theta}
	= \frac{\alpha_e^3}{(1 + \xi)^2} \frac{m^2}{s \, t^2} \Sigma^{\rm LP}_{UU}, \quad
	A^{\rm LP}_{X} = \frac{1}{\Sigma^{\rm LP}_{UU}} \Sigma^{\rm LP}_{X}, \quad
	A^{\rm NLP}_{Y} = \frac{-t}{m \sqrt{\hat{s}}} \frac{1}{\Sigma^{\rm LP}_{UU}} \Sigma^{\rm NLP}_{Y},
\eeq
where $\alpha_e = e^2 / (4\pi)$ is the electromagnetic coupling, $m$ is the nucleon mass, $X \in \{LL, TL\}$ for the LP asymmetries, $Y \in \{UU, LL, (TL,1), (TL,2), UL, LU, (TU,1), (TU,2)\}$ for the NLP asymmetries,
and $\hat{s}$ is the c.m.~energy squared of the $2\to2$ hard scattering between the $A^*$ and $e$, given by 
\beq[eq:sdhep-s-hat-approx]
	\hat{s} = t + \frac{2\xi (s - m^2)}{1 + \xi}
		\simeq \frac{2 \, \xi}{1 + \xi} s + \order{\frac{m^2}{s}, \frac{t}{q_T^2}}.
\eeq
The LP quantities $\Sigma^{\rm LP}_{X}$ are given by 
\bse\label{eq:dvcs-pol-LP}\begin{align}
	\Sigma^{\rm LP}_{UU} &
		= \bb{ \frac{1}{\sin^2(\theta/2)} + \sin^2(\theta/2) } 
			\bb{ \pp{ \frac{1 - \xi^2}{2 \xi^2} \frac{-t}{m^2} - 2 } \pp{ F_1^2 - \frac{t}{4m^2} F_2^2 } 
				- \frac{t}{m^2} (F_1 + F_2)^2 },	\\
	\Sigma^{\rm LP}_{LL} & 
		= \bb{ \frac{1}{\sin^2(\theta/2)} - \sin^2(\theta/2) } 
			(F_1 + F_2) \bb{ F_1 \pp{ \frac{-t}{\xi m^2} - \frac{4 \xi}{1 + \xi} } - \frac{t}{m^2} \, F_2},	\\
	\Sigma^{\rm LP}_{TL} &
		= \frac{\Delta_T}{2m} \bb{ \frac{1}{\sin^2(\theta/2)} - \sin^2(\theta/2) } 
			(F_1 + F_2) \bb{ -4 F_1 + \frac{1 + \xi}{\xi} \, \frac{-t}{m^2} \, F_2 },
\end{align}\ese
which are quadratic in the form factors $F_1$ and $F_2$ defined in \eq{eq:dvcs-a},
\beq
	F_N^{\mu}(p, p') = \vv{ N(p') | J^{\mu}(0) | N(p) }
		= \bar{u}(p') \bb{ F_1(t) \gamma^{\mu} - F_2(t) \frac{i \sigma^{\mu \Delta}}{2 m} } u(p).
\eeq
The momentum $\Delta = p - p'$ is defined as in \eq{eq:two-stage}, with $\Delta_T$ being its transverse component in the diffractive frame,
\beq[eq:delta-T]
	\Delta_T = \frac{\sqrt{(1 - \xi^2) (-t) - 4 \xi^2 m^2}}{1 + \xi}.
\eeq

The NLP quantities $\Sigma^{\rm NLP}_{Y}$ contain both quadratic forms of $(F_1, F_2)$ and linear expressions of the GPD moments,
\beq[eq:gpd-moments]
    V_\F(\xi, t) \equiv \big\{ \H, \, \E, \, \Ht, \, \Et \big\}(\xi, t) 
    = \sum_q e_q^2 \int_{-1}^1 dx \, \frac{\big\{ H^{q, +}, \, E^{q, +}, \, \wt{H}^{q, +}, \, \wt{E}^{q, +} \big\}(x, \xi, t)}{x - \xi + i \epsilon},
\eeq
where we have assembled them in a complex-valued vector $V_\F$, and 
the `$+$' superscripts refer to charge-conjugation-even GPD combinations, 
\beq
    F^{q, +}(x, \xi, t)
    = F^{q}(x, \xi, t) \mp F^{q}(-x, \xi, t),
\eeq
with $\mp$ for $F = H$ or $E$ and $F = \wt{H}$ or $\wt{E}$, respectively.
To write in a compact notation, we introduce the matrix $M$,
\begin{align}
	M = \begin{bmatrix}
		F_1	& -\dfrac{t}{4m^2} F_2		& \xi (F_1 + F_2)	& 0  \\
		(1 + \xi) (F_1 + F_2)	& \xi (F_1 + F_2) 	& \dfrac{1 + \xi}{\xi} F_1	& - \xi F_1 - (1+\xi) \dfrac{t}{4m^2} F_2	\\
		\xi (F_1 + F_2) 	& \pp{ \dfrac{\xi^2}{1 + \xi} + \dfrac{t}{4m^2} } (F_1 + F_2)	
								& - \xi F_1 + \dfrac{t}{4m^2} \dfrac{1 - \xi^2}{\xi} F_2	& - \pp{ \dfrac{\xi^2}{1 + \xi} + \dfrac{t}{4m^2} } F_1 - \dfrac{\xi \, t}{4m^2} F_2 \\
		\xi F_1 - \dfrac{t}{4m^2} \dfrac{1 - \xi^2}{\xi} F_2 	&  \pp{ \xi + \dfrac{t}{4\xi m^2} } F_1 + \dfrac{\xi \, t}{4m^2} F_2
								& -\xi (F_1 + F_2) 	& -\dfrac{\xi \, t}{4m^2} (F_1 + F_2)
	\end{bmatrix}.
\label{eq:coef-M}
\end{align}
Denoting $M_i$ as the $i$-th row vector of $M$, we have the NLP quantities as
\bse\label{eq:dvcs-pol-M}\begin{align}
	\Sigma_{UU}^{\rm NLP} &=
		\frac{\Delta_T}{2m} \frac{1 + \xi}{\xi} 
		\bb{ \frac{2\sin\theta}{\xi} \pp{ F_1^2 - \frac{t}{4m^2} F_2^2 }
			- \frac{4 + (1 - \cos\theta)^2}{\sin\theta \cos^2(\theta/2)} 
				\pp{ M_1 \cdot \Re V_\F }	
		},	\\
	\Sigma_{LL}^{\rm NLP} &=
		- \frac{\Delta_T}{m}
		\bb{ \sin\theta (F_1 + F_2) \pp{ \frac{1 + \xi}{\xi} F_1 + F_2 }
			+ \frac{3 - \cos\theta}{\sin\theta} \pp{ M_2 \cdot \Re V_\F }
		},	\\
	\Sigma_{TL, 1}^{\rm NLP} &=
		2 \sin\theta \, (F_1 + F_2)\bb{ F_1 + \pp{ \frac{\xi}{1 + \xi} + \frac{t}{4\xi m^2} } F_2 }
		+ \frac{2(3 - \cos\theta)}{\sin\theta} \pp{ M_3 \cdot \Re V_\F },	\\
	\Sigma_{TL, 2}^{\rm NLP} &=
	 	2 \sin\theta \, (F_1 + F_2) \pp{ F_1 + \frac{t}{4m^2} F_2 } 
		+ \frac{2(3 - \cos\theta)}{\sin\theta} \pp{ M_4 \cdot \Re V_\F }, \\
	\Sigma_{UL}^{\rm NLP} &=
		- \frac{\Delta_T}{m} \frac{1 + \xi}{\xi} \frac{3 - \cos\theta}{\sin\theta}
		\pp{ M_1 \cdot \Im V_\F }, 
	\\
	\Sigma_{LU}^{\rm NLP} &=
		- \frac{\Delta_T}{2m} \frac{4 + (1 - \cos\theta)^2}{\sin\theta \cos^2(\theta/2)}
		\pp{ M_2 \cdot \Im V_\F },	\\
	\Sigma_{TU, 1}^{\rm NLP} &=
		\frac{4 + (1 - \cos\theta)^2}{\sin\theta \cos^2(\theta/2)}
		\pp{ M_3 \cdot \Im V_\F },	\\
	\Sigma_{TU, 2}^{\rm NLP} &= 
		\frac{4 + (1 - \cos\theta)^2}{\sin\theta \cos^2(\theta/2)}
		\pp{ M_4 \cdot \Im V_\F },
\end{align}\ese

Apparently, the real and imaginary parts of the GPD moments are controlled by the same matrix $M$, 
so the measurement of all these eight NLP polarization asymmetries results in a linear set of equations for the GPD moments,
\beq[eq:F-eqs]
	M \cdot V_\F = \hat{V}_{\rm exp},
\eeq
where $\hat{V}_{\rm exp} = (\hat{V}_{\rm exp}^1, \hat{V}_{\rm exp}^2, \hat{V}_{\rm exp}^3, \hat{V}_{\rm exp}^4)^T$ are the experimentally reconstructed (complex) values of the left-hand sides.
\eq{eq:F-eqs} can be easily inverted to give a unique set of solutions for the GPD moments, $V_\F = M^{-1} \cdot \hat{V}_{\rm exp}$.

\subsection*{B:\ Polarization asymmetry coefficients for the dilepton photoproduction process}
\label{app:tcs}
\vspace{-0.1in}
First, similarly to \eq{eq:dvcs-xsec}, the NLP cross section for the dilepton photoproduction process is
\begin{align}
    \frac{d\sigma^{\ell\ell}_{\gamma}}{dt \, d\xi \, d\phi_S \, d\cos\theta \, d\phi}
    = \frac{d\sigma^{\ell\ell, {\rm unp.}}_{\gamma}}{dt \, d\xi \, d\cos\theta} \, \Omega^{\ell\ell}_{\gamma}(\phi_S, \phi).
\label{eq:tcs-xsec}
\end{align}
We denote the unpolarized cross section $d\sigma_{\gamma}^{\ell\ell, {\rm unp.}}$ 
and the polarization asymmetry coefficients in $\Omega^{\ell\ell}_{\gamma}$ [in \eq{eq:tcs-phi}] as
\beq
	\frac{d\sigma_{\gamma}^{\ell\ell, {\rm unp.}}}{dt \, d\xi \, d\cos\theta}
	= \frac{2\alpha_e^3}{(1 + \xi)^2} \frac{m^2}{s \, t^2} \Sigma^{\rm LP}_{UU}, \quad
	A^{\rm LP}_{X} = \frac{1}{\Sigma^{\rm LP}_{UU}} \Sigma^{\rm LP}_{X}, \quad
	A^{\rm NLP}_{Y} = \frac{-t}{2 m \sqrt{\hat{s}}} \frac{1}{\Sigma^{\rm LP}_{UU}} \Sigma^{\rm NLP}_{Y},
\eeq
where $X \in \{LL, TL, UT\}$ for the LP asymmetries, 
$Y \in \{UU$, $LL$, $(TL,1)$, $(TL,2)$, $UL$, $LU$, $(TU,1)$, $(TU,2)$, $UT$, $LT$, $(TT, 1)$, $(TT, 2)\}$ for the NLP asymmetries,
and $\hat{s}$ is the same as \eq{eq:sdhep-s-hat-approx}.
The LP quantities $\Sigma^{\rm LP}_{X}$ are given by quadratic forms of the form factors $F_1$ and $F_2$,
\bse\label{eq:tcs-pol-LP}\begin{align}
	\Sigma^{\rm LP}_{UU}
		&= \pp{ \frac{1 + \cos^2\theta}{\sin^2\theta} } 
			\bb{ \frac{\Delta_T^2}{2m^2} \pp{ \frac{1 + \xi}{\xi} }^2 \pp{ F_1^2 - \frac{t}{4m^2} F_2^2 } 
				- \frac{t}{m^2} (F_1 + F_2)^2 },	\\
	\Sigma^{\rm LP}_{LL} 
		&= - \pp{ \frac{1 + \cos^2\theta}{\sin^2\theta} } 
			(F_1 + F_2) \bb{ \frac{1 + \xi}{\xi} \frac{\Delta_T^2}{m^2} \, F_1 - \frac{t}{m^2} \, (F_1 + F_2) },	\\
	\Sigma^{\rm LP}_{TL}
		&= \frac{\Delta_T}{2m} \pp{ \frac{1 + \cos^2\theta}{\sin^2\theta} } 
			(F_1 + F_2) \pp{ 4 F_1 + \frac{1 + \xi}{\xi} \, \frac{t}{m^2} \, F_2 }, \\
	\Sigma^{\rm LP}_{UT} 
		&= - \frac{\Delta_T^2}{2m^2} \pp{ \frac{1 + \xi}{\xi} }^2 \pp{ F_1^2 - \frac{t}{4m^2} F_2^2 },
\end{align}\ese
where $\Delta_T$ is the same as \eq{eq:delta-T}.
As in \eq{eq:dvcs-pol-M}, we write these equations in a compact form showing the linear dependence of the NLP quantities $\Sigma^{\rm NLP}_{Y}$ on the GPD moments, 
by introducing an additional matrix $\wt{M}$ that differs from $M$ only by flipping the signs of the last two columns.
Since the GPD moments in the dilepton photoproduction differ from the photon electroproduction only by a complex conjugate, 
we still use $V_\F$ in \eq{eq:gpd-moments} and write $\Sigma^{\rm NLP}_{Y}$ as
\bse\label{eq:tcs-pol-M}\begin{align}
	\Sigma_{UU}^{\rm NLP} &=
		\frac{\Delta_T}{m} \frac{1 + \xi}{\xi} 
		\bb{ - \frac{4\cot\theta}{\xi} \pp{ F_1^2 - \frac{t}{4m^2} F_2^2 }	
			+ \frac{1 + \cos^2\theta}{\sin\theta} \pp{ M_1 \cdot \Re V^*_\F }	
		},	\\
	\Sigma_{LL}^{\rm NLP} &=
		\frac{\Delta_T}{m}
		\bb{ \frac{4 \cot\theta}{\xi} (F_1 + F_2) \pp{ (1+\xi)F_1 + \xi F_2 }
			- \frac{1 + \cos^2\theta}{\sin\theta} \pp{ M_2 \cdot \Re V^*_\F }
		},	\\
	\Sigma_{TL, 1}^{\rm NLP} &=
		- 8(F_1 + F_2) \cot\theta \bb{ F_1 + \pp{ \frac{\xi}{1 + \xi} + \frac{t}{4\xi m^2} } F_2 }
		+ 2 \frac{1 + \cos^2\theta}{\sin\theta} \pp{ M_3 \cdot \Re V^*_\F },	\\
	\Sigma_{TL, 2}^{\rm NLP} &=
	 	- 8 \cot\theta \, (F_1 + F_2) \pp{ F_1 + \frac{t}{4m^2} F_2 } 
		+ 2  \frac{1 + \cos^2\theta}{\sin\theta}
		\pp{ M_4 \cdot \Re V^*_\F },	\\
	\Sigma_{UL}^{\rm NLP} &=
		- \frac{\Delta_T}{m} \frac{1 + \xi}{\xi} \frac{1 + \cos^2\theta}{\sin\theta}
		\pp{ M_1 \cdot \Im V^*_\F }, \\
	\Sigma_{LU}^{\rm NLP} &=
		\frac{\Delta_T}{m} \frac{1 + \cos^2\theta}{\sin\theta}
		\pp{ M_2 \cdot \Im V^*_\F },	\\
	\Sigma_{TU, 1}^{\rm NLP} &=
		-2 \frac{1 + \cos^2\theta}{\sin\theta}
		\pp{ M_3 \cdot \Im V^*_\F },	\\
	\Sigma_{TU, 2}^{\rm NLP} &= 
		2 \frac{1 + \cos^2\theta}{\sin\theta}
		\pp{ M_4 \cdot \Im V^*_\F },	\\
	\Sigma_{UT}^{\rm NLP} &=
		- \frac{\Delta_T}{m} \frac{1 + \xi}{\xi}
		\bb{ \frac{4 \cot\theta	}{\xi} \pp{ F_1^2 - \frac{t}{4m^2} F_2^2 }
			+ \sin\theta \, \bigpp{ \wt{M}_1 \cdot \Re V^*_\F }
		}, 
	\\
	\Sigma_{LT}^{\rm NLP} &=
		- \frac{\Delta_T}{m} \sin\theta \,
		\bigpp{ \wt{M}_2 \cdot \Im V^*_\F }, 
	\\
\Sigma_{TT, 1}^{\rm NLP} &=
		2 \sin\theta \, \bigpp{ \wt{M}_3 \cdot \Im V^*_\F },	\\
\Sigma_{TT, 2}^{\rm NLP} &=
		-2 \sin\theta \, \bigpp{ \wt{M}_4 \cdot \Im V^*_\F },
\end{align}\ese
where $\wt{M}_i$ denotes the $i$-th row of $\wt{M}$.

Compared to the photon electroproduction results in \eq{eq:dvcs-pol-M}, we now have four more constraints on the GPD moments from 
$(A_{UT}^{\rm NLP}$, $A_{LT}^{\rm NLP}$, $A_{TT, 1}^{\rm NLP}$, $A_{TT,2}^{\rm NLP})$.
They are not independent from the other eight. 
Specifically, $\wt{M}_1$ can be written as a linear combination of $M_1$, $M_2$, and $M_3$, so its information on the real parts of the GPD moments is
covered by $(A_{UU}^{\rm NLP}, A_{LL}^{\rm NLP}, A_{TL, 1}^{\rm NLP})$.
Similarly, $\wt{M}_2$, $\wt{M}_3$, and $\wt{M}_4$ can be written as linear combinations of $(M_1, M_2, M_4)$, $(M_1, M_3, M_4)$, and $(M_2, M_3, M_4)$, respectively.
Nevertheless, the linear photon polarization might be more easily controlled in experiments than the transverse target spin,
so the asymmetry $A_{UT}^{\rm NLP}$ can be used in place of $A_{TL, 1}^{\rm NLP}$, 
and $A_{LT}^{\rm NLP}$ in place of $A_{TU, 2}^{\rm NLP}$.

\end{document}